\documentclass{PoS}

\title{Disorder parameter for confinement and vacuum field strength correlators.}

\ShortTitle{Disorder parameter for confinement.}

\author{\speaker{Adriano Di Giacomo}\\
      \\ Pisa University and INFN
      
        E-mail: \email{adriano.digiacomo@df.unipi.it}}

\abstract{Abstract.\\The possibility is explored to relate confinement to properties of gauge invariant field strength correlators.
         }

\FullConference{The XXV International Symposium on Lattice Field Theory\\
		 July 30-4 August 2007\\
		 Regensburg, Germany}

\begin{document}

\section{Introduction}
 A disorder parameter for dual superconductivity of gauge theory vacuum has been developed \cite{DigPaf} \cite{DigLuc1}\cite{DigLuc2}\cite{Frolich}. It is the $vev $ , $\langle \mu \rangle$ , of an operator carrying non zero magnetic charge ${q\over {2g}}$.
 The Euclidean version is 
 \begin{equation}
 \mu(\vec x, t) = \exp [{ -{q\over {2g^2}}\int d^3y \vec E(\vec y, t) \vec b_{\perp}(\vec x -\vec y)] }
 \end{equation}
 Here       $\vec b_{\perp}(\vec r) = {{\vec n \wedge \vec r}\over {r(r -\vec r.\vec n}} $         is the field of a monopole in the transverse gauge with
 
 $\nabla \vec b_{\perp}=0$  
 
  $\nabla \wedge \vec b_{\perp}={\vec r \over r^3} - 4\pi \theta(\vec n.\vec r)\vec n \delta^2(\vec r_{\perp})$.
  
  The field  $\vec E_{\perp} $ is the conjugate momentum to the transverse component of the potential $\vec A_{\perp}$ so that $\mu$ is nothing but a translation operator of $\vec A_{\perp}$ , or
  \begin{equation}
  \mu(\vec x, t) | \vec A_{\perp}(\vec y,t)\rangle = | \vec A_{\perp}(\vec y,t) + {q\over {2g}} \vec b_{\perp}(\vec x - \vec y) \rangle
  \end{equation}
  It just creates a monopole.
  
   One of the  factors  ${1\over g}$ at the exponent of  Eq(1.1) comes from the Dirac quantization condition for magnetic charge, the other one from the fact that  the electric field in the lattice formulation has an additional multiplicative factor  $g$ .  The operator $\mu$ can be written  in the form
  \begin{equation}
  \mu  =  \exp{[ -\beta \Delta S]}
  \end{equation} 
  with the usual notation $\beta = {{2N}\over {g^2}}$ and  $\Delta S = -{q\over{4N}} \int d^3y \vec E (\vec y, t) \vec b_{\perp}(\vec x -\vec y)$.
  As a consequence   $ \langle \mu \rangle$ is the ratio of two partition functions  and  $\langle \mu \rangle =1 $ at $\beta = 0$.
  \begin{equation}
  \langle \mu \rangle  = { {Z(S + \Delta S)}\over {Z(S)}}
  \end{equation}
  
  For compact $U(1)$ gauge theory  a few theorems have been proved:
  
  (1)  $  \mu(\vec x, t) $  is a gauge invariant, Dirac like , magnetically charged operator , and obeys cluster property\cite{Frolich}\cite{Cirigliano}.
  
  (2) $ \langle \mu  \rangle  \neq 0 $ for $\beta < \beta_c$ where there is confinement ,  $ \langle \mu  \rangle = 0$ for $\beta \ge \beta_c$ i.e. in the deconfined phase. $\beta_c $ is the critical  point.
  
  In $U(1)$ gauge theory   confinement is therefore produced by condensation of monopoles i.e.  by dual superconductivity of the vacuum.
  
  Instead of $\langle \mu \rangle$ it proves convenient  to use the quantity 
  \begin{equation}
   \rho ={{\partial \ln(\langle \mu \rangle )}\over {\partial \beta}}
   \end{equation}
    From Eq(1.4) it follows that  $\rho =\langle S \rangle_S - \langle  S + \Delta S \rangle_{S + \Delta S}$  , the subscript indicating the action used in the statistical weight. Moreover because of the boundary condition $\mu =1$ at $\beta=0$,
  
  \begin{equation}
  \langle \mu  \rangle = \exp [\int_0^{\beta}\rho(\beta') d\beta']
  \end{equation}
  For $SU(N)$ gauge theories, with and without quarks, $N-1$ operators   $ \mu^a $   can be defined $(a=1,2,...N-1)$, and the corresponding order parameters  $\langle \mu^a \rangle$\cite{DigLuc1},\cite{DigLuc2}.  The definition of  $ \mu^a $  has the same form as $\mu$ in Eq (1.1) with the field strength $\vec E_{\perp}$  replaced by  $\vec E_{\perp}^a(\vec y, t) \equiv Tr [ \Phi^a \vec E(\vec y, t)]_{\perp}$  where  $\Phi^a $  selects the direction of the residual abelian gauge field in the abelian projected gauge. For a detailed discussion
  see Ref's\cite{DigLuc1},\cite{DigLuc2}. The choice of the abelian projection is irrelevant \cite{CarDig1}\cite{Digp}\cite{DigPaf2}.
  
  Fig(1) shows the numerical determination of   $ \langle \mu  \rangle  $ for  compact $U(1)$ gauge theory, Fig(2) shows the corresponding quantity $\rho$ , which presents a strong negative peak  at the critical point $\beta _c$ .  The analysis goes as follows\cite{DigLuc1},\cite{DigLuc2}:
   \begin{figure}[ht]
   \begin{center}
\includegraphics[width=.5\textwidth,clip=]{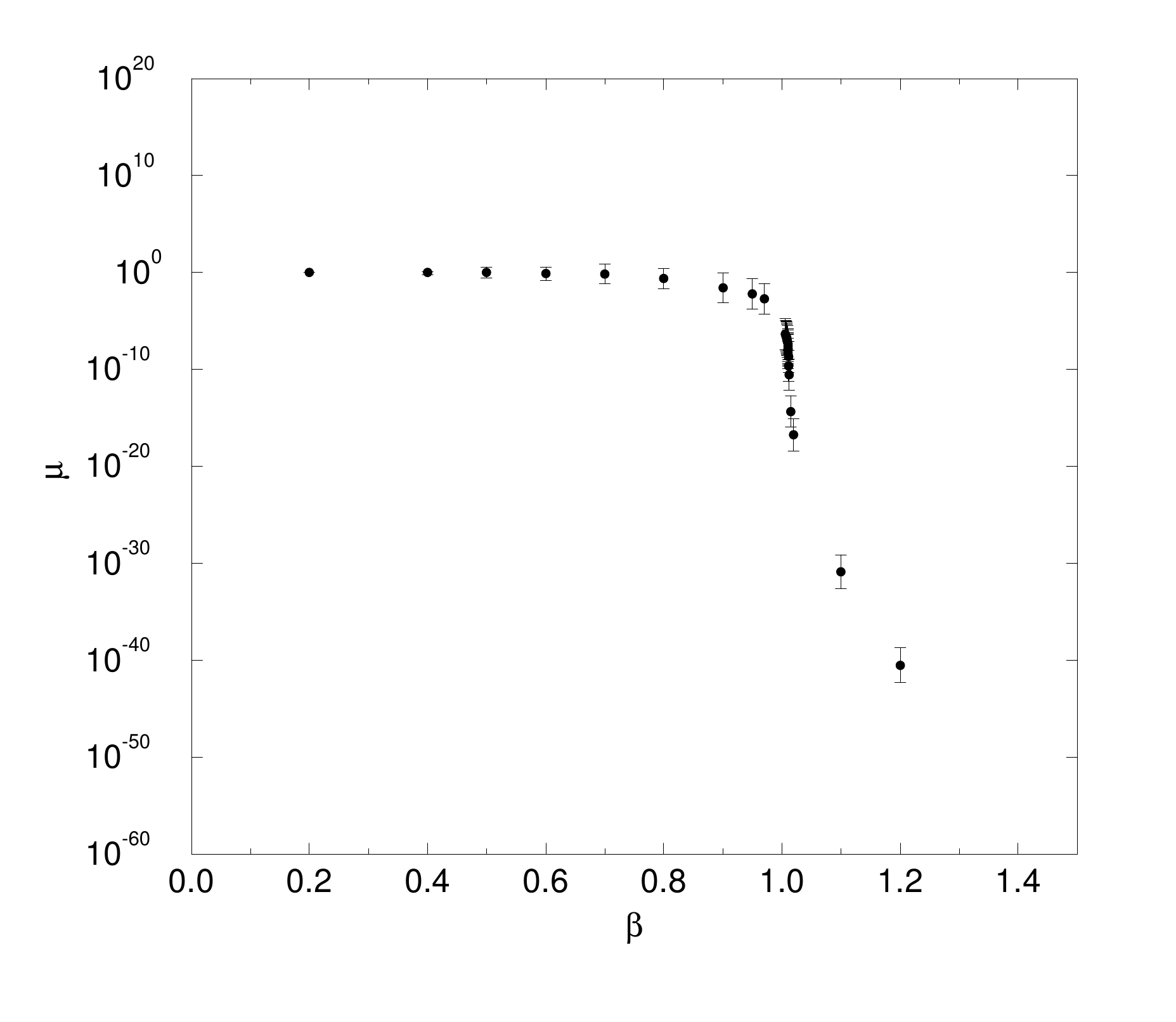}
 \caption{ $\langle \mu \rangle$ for U(1) lattice gauge theory  Ref\cite {DigKyo}.
}
\end{center}
\end{figure}
\begin{figure}
\begin{center}
\includegraphics[height=.25\textheight]{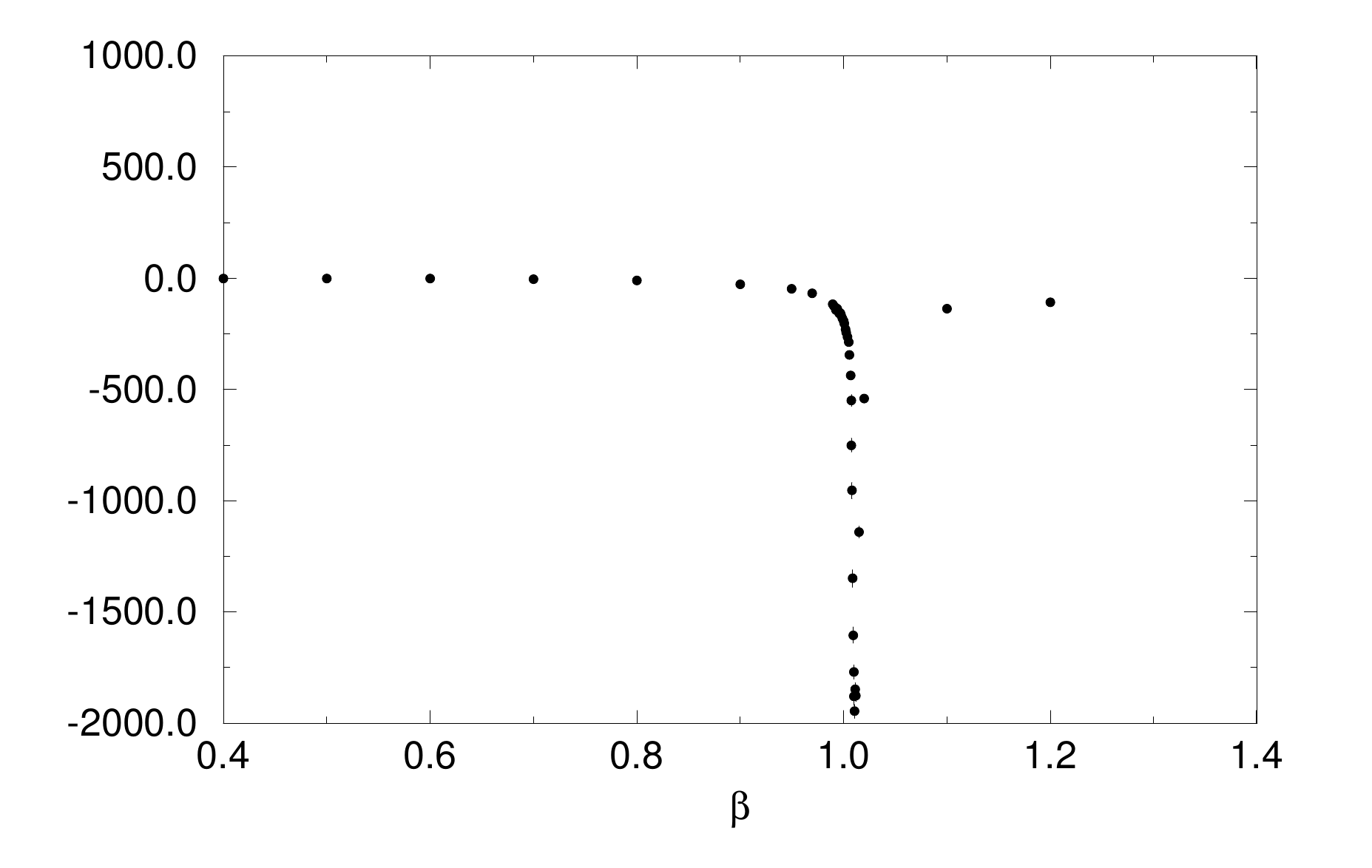}
\caption{ $\rho$ versus $\beta$  Ref.\cite{DigPaf}}.
\end{center}
\end{figure}

  1)  For $\beta < \beta_c$ $\rho^a $ tends to a finite limit in the thermodynamical limit $V=L_s^3 \to \infty$ , and by use of Eq(1.6) $\langle \mu \rangle \neq 0$
  
  2) For $\beta \ge \beta_c$         $\rho^a \approx - |c| L_s + c' $  with $c \neq 0$ ,or , by use of Eq(1.6)  $lim_{L_s \to\infty} \langle \mu  \rangle  =0 $
  
  3)At $\beta \approx \beta_c$ the correlation length goes large compared to the lattice spacing and the scaling law holds
  \begin{equation}
  {{\rho}\over {L_s^{1\over \nu}} }  =  f(\tau L_s^{1\over \nu}) 
  \end{equation}
    where  $\tau \equiv 1 - {T\over T_c} $.  $\nu$ is the critical index of the correlation length , and is typical of the universality    
    class of the transition. For weak first order  ${1\over \nu }=  3$ [ Quenched $SU(3)$\cite{DigLuc2} , $N_f=2$  $QCD$ \cite{DelDigPic}], for 3d-Ising  ${1\over \nu} = 1.6$  [ SU(2) \cite 
    {DigLuc1}], for $3d- O(4)$ ${1\over \nu} =  1.336$.

  The properties 1), 2), 3) as observed in different systems are shown in Figs (3), (4), (5)
  \begin{figure}
  \begin{center}
\includegraphics[height=.25\textheight]{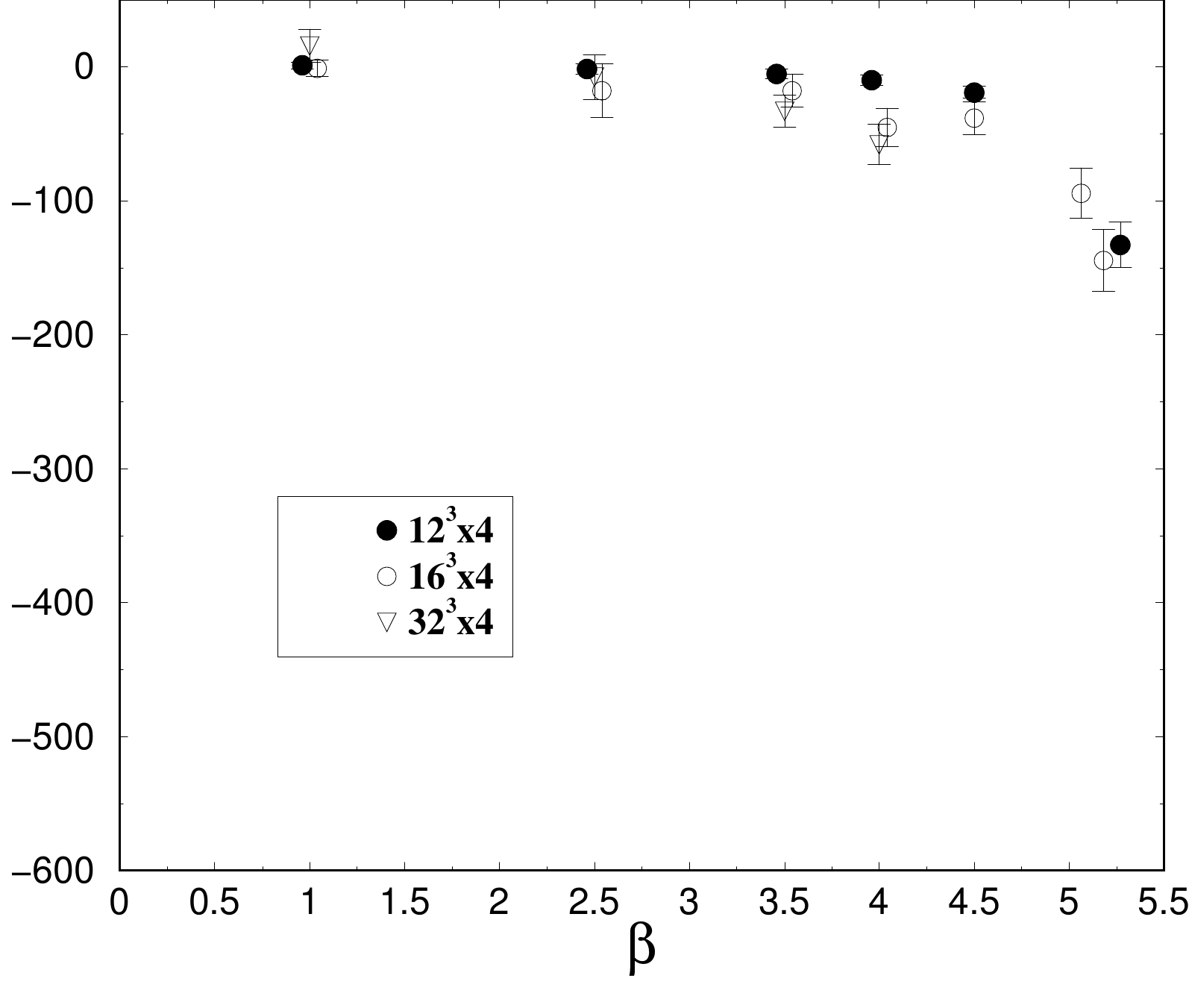}
 \caption{ Strong coupling behavior of $\rho$ at various lattice sizes and $am=0.1335$ Ref.\cite{CarDig2}}
 \end{center}
\end{figure}
\begin{figure}
\begin{center}
\includegraphics[height=.25\textheight]{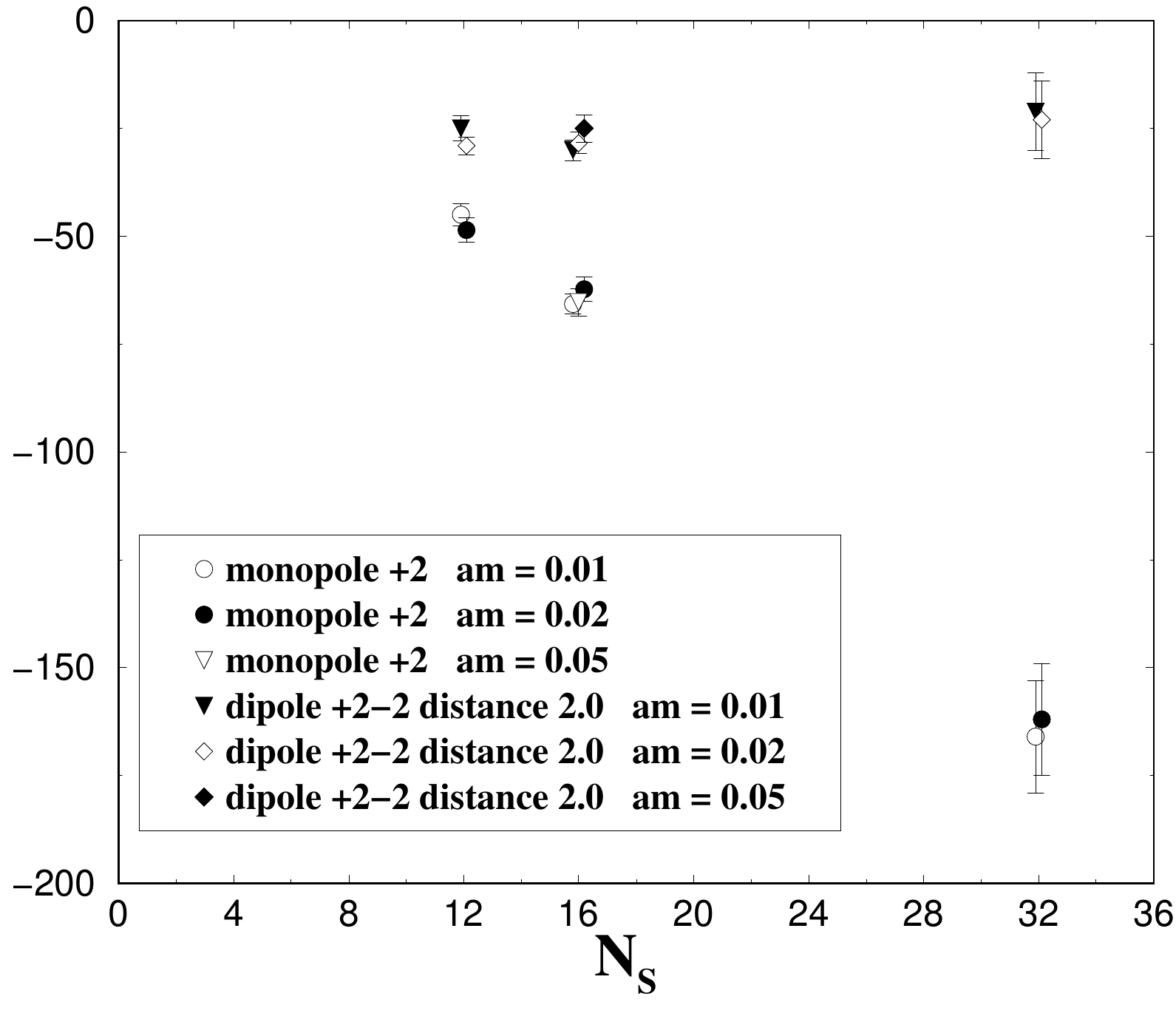}
 \caption{  Volume dependence of $\rho$ in the deconfined phase for different values of the magnetic charge.Ref \cite{DelDigLuc}}
 \end{center}
\end{figure}
\begin{figure}
\begin{center}
\includegraphics[height=.25\textheight]{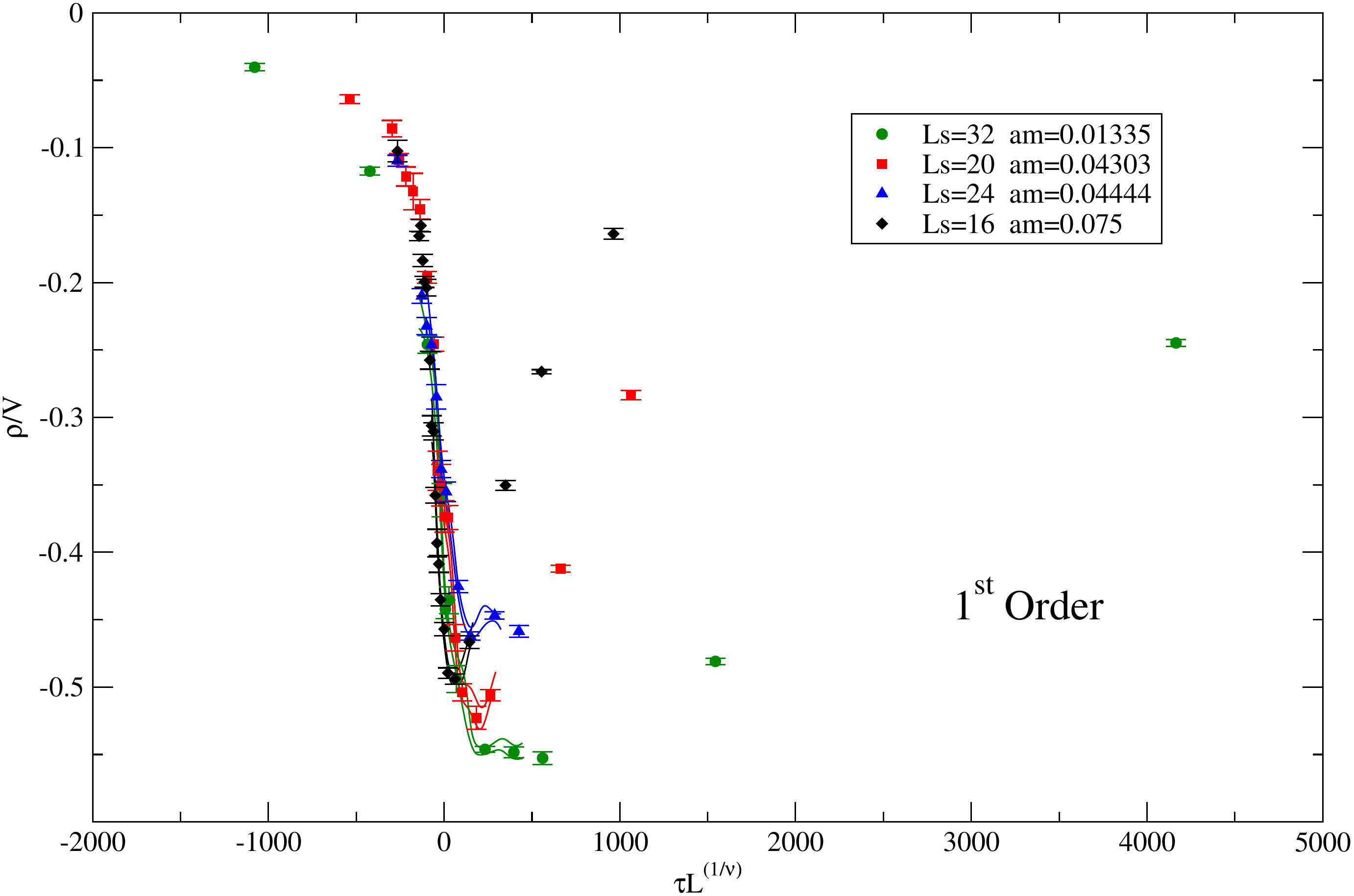}
 \caption{ Scaling of  $\rho$ assuming first order for the deconfining transition. $N_f=2 $ $QCD$. Ref. \cite{DelDigPic}}
 \end{center}
\end{figure}

  The question we address \cite{Dig} in this paper is whether  $\langle \mu  \rangle $ can be computed in the frame of the Stochastic Vacuum Model of QCD\cite{Dos}\cite{Sim} \cite{DigDosSimShe}
 . The model consists in expressing physical quantities in terms of gauge invariant correlators of field strengths , making a cluster expansion of them and keeping  only the two point cluster.
 The model provides a good description of many aspects of $QCD$ and it would be interesting to know if the distinction between confined and deconfined phase could be read in the behavior of the correlators.
 
 \section{Cluster expansion of  $\langle \mu^a  \rangle$ }
 
 The series expansion of the exponential in Eq(1.1) reads \cite{Dig}
\begin{eqnarray}
 &&\langle    \exp[ -{q\over {2g^2}}\int d^3y \vec E^a(\vec y, t) \vec b_{\perp}(\vec x -\vec y) ]    \rangle =  \\  &&\Sigma_n{1\over n!}(-{q\over {2g}})^n \int d^3x_1 \int d^3x_2...\int  d^3 x_n \vec b^{i_1}_{\perp}(\vec x - \vec y_1) .......\vec b^{i_n}_{\perp}(\vec x - \vec y_n)  \nonumber
   \langle E^a_{i_1}(\vec y_1, t)..... E^a_{i_n}(\vec y_n , t) \rangle
 \end{eqnarray}
 According to the stochastic vacuum model one performs a cluster expansion of the correlators: the one point cluster is zero by symmetry, and clusters of order higher than 2 are neglected.  Keeping the correct combinatorics into account \cite{Dig} the net result is 
 \begin{equation}
 \langle \mu \rangle  =  \exp [ - {q^2\over {8g^2}}\int d^3y_1 d^3y_2  b_{\perp}^{i_1}(\vec y_1 - \vec x) b_{\perp}^{i_2}(\vec y_2 - \vec x)\langle E^a_{i_1}(\vec y_1 ,t) E^a_{i_2}(\vec y_2 ,t)\rangle]
 \end{equation}
 Higher clusters are $O(q^4)$ at the exponent. Here $\vec E^a = Tr[ \Phi ^a \vec E] $. 
 The gauge invariant correlator at the exponent   of Eq(2.2) 
 \begin{equation}
 \langle E^a_{i_1}(\vec y_1 ,t) E^a_{i_2}(\vec y_2, t) \rangle = \Phi^a_{i_1 i_2}( \vec y_1 - \vec y_2)
 \end{equation}
 in principle depends on the path $C$ used to parallel transport from $\vec y_1$ to $\vec y_2$  but this dependence is irrelevant to the study of the ultraviolet and infrared behavior.
 Since $\beta \equiv {{2N}\over g^2} $ , by use of Eq's (1.5), (2.2) we get
 \begin{equation}
 \rho ^a = {\partial \over {\partial \beta}}[-{q^2 \over {16N}}\beta  \int d^3 y_1 \int d^3 y_2 b_{\perp}^{i_1} (\vec y_1) b_{\perp}^{i_2} (\vec y_2)\Phi^a_{i_1 i_2}( \vec y_1 - \vec y_2)]
 \end{equation}
 In Fourier transform
 
\begin{equation}
\vec b_{\perp}(\vec k) = {{\vec n\wedge \vec k}\over {k(k - \vec k .\vec n +i\epsilon)}}    ,    \vec H (\vec k) =  -i [{{\vec k}\over k^2} - {{\vec n}\over {(\vec n\vec k -i\epsilon})}]
\end{equation}
\begin{equation}
\Phi^a_{ij} (\vec k)  =  (k^2 \delta_{ij} - K_i k_j) f(k^2)
\end{equation}
 independent on $a$ \cite{DigDosSimShe}.
\begin{equation}
(k^2 \delta_{ij}  -  k_i k_j) b^i_{\perp}(\vec k)  b^j_{\perp}(- \vec k) = |\vec H(\vec k)|^2  = {1\over {k_z^2}} - {1\over k^2}
\end{equation}
The equation follows for $\rho$
\begin{equation} 
\rho = - {q^2\over{16N}} {\partial\over {\partial \beta}}[\beta \int {{d^3k}\over{(2\pi)^3}} f(k^2)( {1\over {k_z^2}} - {1\over k^2})]
\end{equation}
In the deconfined phase the perturbative expression can be used       $ {f(k^2) \over {(2\pi)^3}} = {1\over {2k}}$ 
and
\begin{equation} 
\rho^a = {{\pi q^2}\over {8N}}[-\sqrt{2} L_s + 2 \ln(L_s) + constant]
\end{equation}
with $L_s$ the spatial size of the lattice , and
\begin{equation}
L_s  =  {{(L_s a)}\over a} \equiv { {IR-cutoff }\over {UV-cutoff}}
\end{equation}
 In the thermodynamical limit  $L_s \to \infty $ ,$\rho \to -\infty$  as in Fig(4), and by use of Eq(1.5), $\mu \to 0$.
 
 In the confined phase one expects the same $UV$ behavior, which is dictated by $OPE$ ,but  $a L_s$ will be replaced by some $IR$ cutoff
 $\Lambda$ so that ,at fixed lattice spacing $\rho^a$ is volume independent as in Fig(3) . 
 
 This will never be the case if Eq(2.8) holds, no matter how $IR$
 well behaved is the correlator: the term $\propto {1\over {k^2_z}}$ due to the Dirac string will always diverge.
 
 This means that the stochastic approach is inadequate in the confined phase. Indeed in that phase the vacuum is a Bogolubov-Valatin superposition of states with different magnetic charge  and the operator $\mu$ will connect sectors differing by  $q$ units of magnetic charge : The Dirac string will then end on an antimonopole and the integral will be $IR$ cut-off by a massive propagator and be volume independent.
 
 This can be checked in $U(1)$ theory in the dual formulation of Polyakov \cite{Pol}.The potential of the dual field $\chi$ is there proportional to $cos(\chi)$
 which in the weak coupling is equivalent to a mass term, and gives a gaussian distribution like  the stochastic vacuum model in $QCD$. In the strong coupling regime the tunneling between the minima of $cos(\chi)$ provides a Bogolubov-Valatin vacuum.
 
  Like the Polyakov line the order parameter $\langle \mu \rangle$ is singular in the continuum limit $a \to 0$,
 but its $IR$ behavior at any finite $UV$ cutoff detects confinement or deconfinement.
 \begin{figure}
 \begin{center}
 \includegraphics[height=.25\textheight]{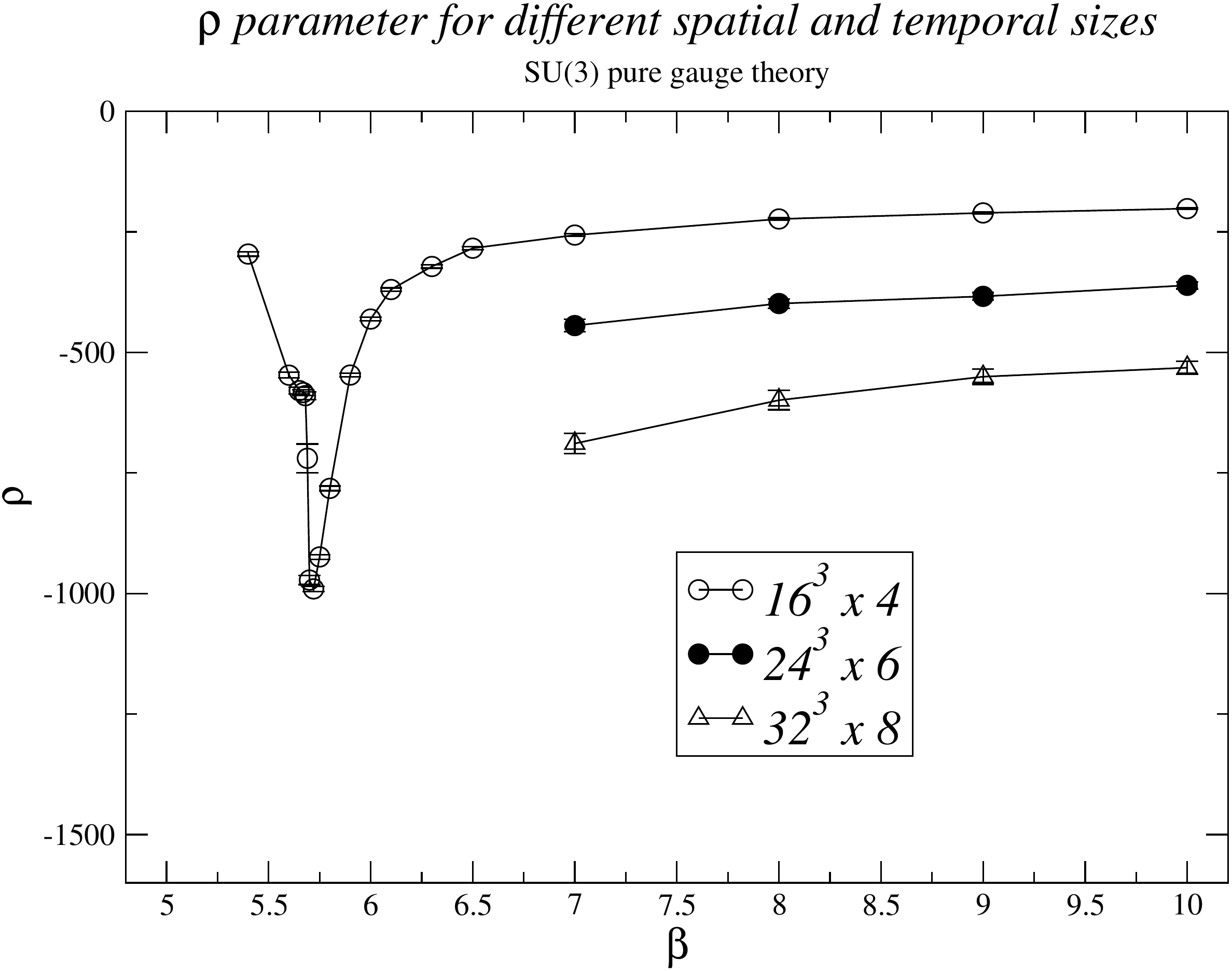}
 \caption{ Check of Eq(2.9)}
 \end{center}
 \end{figure}


\end{document}